# HOME HEALTHCARE PROCESS: CHALLENGES AND OPEN ISSUES

Selma ARBAOUI[1], Nathalie CISLO[2], Natalie SMITH-GUERIN[3]

[1]Selma.Arbaoui@univ-orleans.fr, [2]Nathalie.Cislo@bourges.univ-orleans.fr,

[3]Natalie.Smith@bourges.univ-orleans.fr

Université d'Orléans, Institut PRISME, 63, Av de Lattre de Tassigny, 18020, Bourges Cedex, France.

**Abstract:** *Home healthcare is part of the most critical research and development healthcare areas. The objective is to decentralize healthcare, leading to a shift from in-hospital care to more advanced home healthcare, while improving efficiency, individualisation, equity and quality of healthcare delivery and limiting financial resources. In this paper, we adopt a process approach to tackle the home healthcare domain in order to highlight the importance of organisational aspects in the success of an ICT-home healthcare project. Such projects should be supported by an automated system, called in this paper, Home Healthcare support system. We examine HH processes from two selected perspectives (complexity and dynamics) to illustrate the requirements of a HH support system. We advocate that satisfying these requirements is part of the most important challenges in the home healthcare research domain and we propose first track of solutions by attempting to benefit from past experiences in 3 process research communities.*

## 1. Introduction

Nowadays, developed and industrialized countries are facing important problems regarding healthcare services, such as: increase in the number of elderly people and in chronic diseases, demand for accessibility of care outside hospitals (into patients' own homes in order to enable them to spend more time in a familiar environment), the need to reduce hospitalisation costs... In parallel the countries have to improve efficiency, individualisation, equity and quality of healthcare delivery while limiting financial resources. So with the increasing prevalence of chronic diseases and the ageing of the population, the focus of healthcare "*expanded from adding years to life to adding life to years*" (Stefanelli 2002). For all these reasons, home healthcare is part of the most critical research and development healthcare areas. The objective is to decentralize healthcare, leading to a shift from in-hospital care to more advanced home healthcare.

Home healthcare is a very broad term and may comprise everything from *basic home healthcare* to *advanced home healthcare*. The term *basic home healthcare* is used when health services are rendered in the home to the aged or disabled individuals. Different persons and services are implied in such case: medical and para-medical professionals, nursing services, physical, homemaker services, social services. At the patient home, family members are also implied in the healthcare delivery. *Advanced Home Healthcare* is used in place of *ICT-Based Home healthcare*. It may comprise technological solutions as e-mail consultations, advanced sensor surveillance (for collecting vital signs and physiological parameters), clinical robots, etc. The patient may be less frequently visited than in regular *basic home healthcare*.

In this paper *Home Healthcare* (HH) refers to *advanced home healthcare* and the focus is put on the home healthcare process that we consider as one of the *"human intensive processes"*. We adopt a process approach to tackle the home healthcare domain in order to highlight the importance of organisational aspects in the success of an *ICT-home healthcare project*. The stated objectives of such a project are:

- implementing a given technology in a home healthcare organisation (e.g. a telemedicine solution),
- defining and managing the resulted ICT-based home healthcare process.

We suggest that such projects should be supported by an automated system, called in this paper, *HH support system*. By "supporting" we mean to provide capabilities for mastering the HH process. During the HH project, the *HH support system* should provide capabilities for analysing, modelling, simulating, enacting (*Enactment* means the manual or automatic execution of an instantiated process model. Enacting a process model may result in: guiding, assisting and why not automating, when possible, this process, i.e. the home healthcare activities), and improving the HH process.

In this paper, we examine HH processes from two selected perspectives (complexity and dynamics) to illustrate the requirements of a *HH support system*. We advocate that satisfying these requirements is part of the most important challenges in the HH research domain and we propose first track of solutions by attempting to benefit from past experiences in 3 process research communities.

The investigated process communities are respectively: manufacturing process engineering, software process engineering and business process engineering. Processes have been differently tackled according to concerns, background and objectives of each community. Existing differences are motivated for example by the finality of process modelling (to enact? to simulate? to analyse? to improve or to re-engineer?) or the handled perspective (function, information, process, organization, operation).

The remainder of this paper is organized as follows. Section 2 presents the benefits of a process approach for supporting an *ICT-Based home healthcare project* and introduces the HH process along with its features. Next, Section 3 presents the selected requirements that have to be met by a *HH support system* and section 4 the possible tracks of solutions derived from manufacturing, software and business process communities results. Open issues and perspective works are given in section 5.

**2. The Home Healthcare process**

*Why a process approach?*
To tackle the home healthcare domain, we have adopted a process approach to highlight the importance of the organisational aspects in the success of an *ICT-home healthcare project*. Indeed, existing research works on home healthcare have dealt more with developing technologies for home telehealth, home monitoring, or home telemedicine, leaving process aspects questions unanswered (Koch 2005).

In addition, some recent observations, reported by (Beuscart et al. 2004), indicate that the requirements of the homecare actors (nurses, physicians, home healthcare organisations, patients families) are more oriented towards the improvement of the organisation and management of the homecare system over a more intensive use of home telemedecine.

Besides, a recent evaluation work of some telemedicine projects shows that ICT tools are not "care producers", but they improve home care provision if they lay on controlled processes, which are well known, identified, coherent and validated ( (Telemedicine 2000), http://www.sante.gouv.fr/htm/dossiers/telemed/eval/).

These studies and observations highlight the importance of organisational aspects in an *ICT-home healthcare project*, not only during the process care but also upstream and downstream:

**Upstream:**
- It is necessary to clearly identify and analyse the process in order to detect possible dysfunctions and decide if technological support may improve the process care and if it can improve it: which kind of technology may be used? For example, in some cases the problem may be solved by reorganising home visits instead of adopting a telesurveillance solution.

**Downstream:**
- Mastering the processes may help in answering questions like: how can we take advantage of home healthcare technology and how can we organize the delivery of healthcare?

The mastering of the HH process is a necessary condition to provide an efficient support for such home healthcare project (Upstream and Downstream). As suggested before, these capabilities may be provided by a *HH support system*. But before going on about this system, let us specify in the following what is a "Home Healthcare Process" (according to us) and what is the nature of this process?

After investigating some local home-care services in addition to those described in some papers (Arbaoui et al. 2007, Bricon et al. 2005), we consider that a home healthcare process may encapsulate:

- An **organisational process** made up of activities at a strategic level, such as: managing the home care organisation that produces the care (from the private or public sector), contracting new patients, managing the logistical aspects of the care activities and human resources. This may include managing materials suppliers (beds, drugs, telesurveillance tools, sensors, etc.), managing the network of the different actors involved in the care process: medical, para-medical, social, etc. and managing funding aspects (Insurances, Social security, etc.)
- An **organisational care process** made up of activities at tactic and operational levels, such as: patient admissions, evaluation and planning, elaboration and validation of a care project (therapeutic, educative, social), the organisation of care activities following an alert (triggered by a sensor placed in the patient's home), home visit planning and monitoring, managing the discharge of the patients.
- A **care process** made up of day-to-day care activities delivered at home by nurses, doctors or the family's patient members (for simple procedures).

In this paper, the focus is put specifically on the tactic and operational aspects**,** thus the HHP (or HH process) refers to **the organisational care process**.

Concerning this process features, we consider the HHP as one of the "human intensive processes". Such processes are « *usually long lived, distributed among various participants, made of heterogeneous components with a various level of autonomy and always subject to dynamic evolution* » (Cunin 2000).

The consequences are the following features; some of them are common to both "traditional" and home healthcare. However, we noticed several differences:

- Home healthcare processes are very complex. They involve clinical and administrative tasks. The persons in charge of patients admission (coordinating nurses in most cases) are also in charge of the clinical procedure, i.e. evaluation of the clinical situation of the patient, they are also in charge of the administrative procedure, they try to facilitate the patient's home care by providing (or help families to get) specific healthcare materials, to arrange home patients with limited mobility, to organize (or help families to do so) their meals, house work, etc. "Traditional" healthcare processes are not concerned by these activities as the patient is entirely taken care of by the hospital.

- They integrate heterogeneous sub-services and involve multiple private or public organisations: healthcare organisations, social organisations, home-care materials and technologies suppliers, chemists, patients transportation. During traditional hospitalization, most of these services are provided by the hospital employees.

- Home Healthcare processes are also very dynamic; this is one of their main intrinsic characteristics. This means that they are continually changed, adapted and completed. The major reasons are:
    - They often cannot be completely defined in advance.
    - People often have to cope with unforeseen situations (arising from human or material issues).

- Variability of the patient environment.

Finally, HH processes may run over a long period of time, as for instance in the case of chronic diseases where management over time is essential (Holman and Lorig 2004). Therefore, these processes have to be frequently adapted (due to new laws, new medical treatments, operational reorganisations ...).

- Home healthcare activities are decentralized, involving different actors (e.g. the family doctor, nurses, family members …) whose cooperation and coordination are absolutely required to achieve high quality home healthcare. The traditional single doctor-patient relationship is being replaced by a new one in which the patient is managed by a team of healthcare professionals, each specialized in one aspect of the care and where common meetings between actors rarely occur.

Although it is an important characteristic of the home healthcare, this last feature, will not be tackled in this paper which focus on the process complexity and dynamics (i.e. agility, flexibility). These two features lead us to two main requirements that have to be fulfilled by a *HH support system*. As related before, answering to these requirements should be critical research issues in the home healthcare domain and will represent real challenges in a near future. In what follows we will firstly present these requirements using the HH process framework and we will try to answer to each of them by attempting to benefit from past experiences in 3 process research communities.

**3. Home Healthcare support system: Which requirements?**

In this section the requirements induced by HH process complexity and dynamics are presented using the HH process framework (Inspired by Dowson's Framework (Dowson 1992)) (see figure 1) that consists in three domains: modelling, enactment, day to day home healthcare process (day to day activities). This framework will allow us to determine precisely to which domain (modelling, enactment,...) a given requirement is related and thus to precisely determine how to meet it and what will be the possible consequences on the *HH support system* if the requirement is not met:

*Home healthcare process domain:*
Involves tasks and activities that are performed by human/non human home healthcare actors (doctors, nurses, sensors, web-based technologies, video, etc…). These tasks/activities may take place at different level of granularity like for instance "scheduling nurses home visits", "managing a patient admission or discharge" (by coordinators personal that are often nurses), "performing a blood test at home" (by a nurse), "giving medicine" (by a member of the patient's family), "monitoring vital sign parameters (VSP)" (by a sensor), etc.

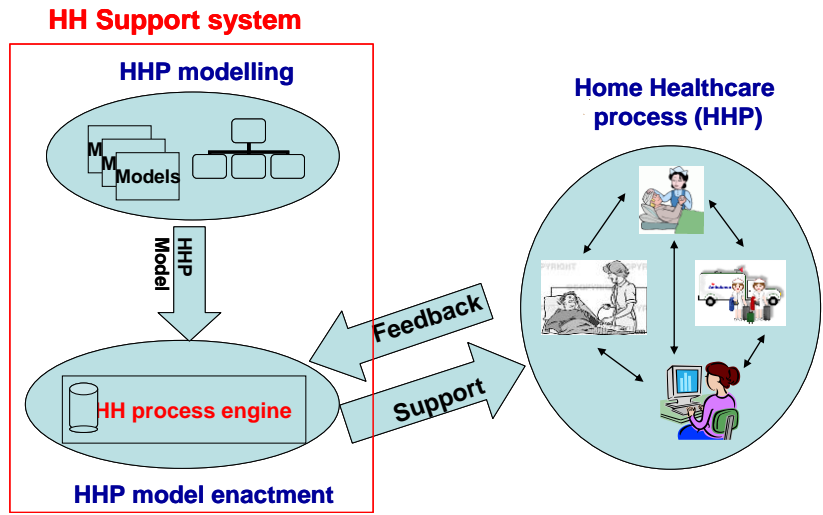

**Figure 1: HH process framework**

*Home healthcare process modelling domain:*
Contains specifications of home healthcare processes (or fragments of these processes), expressed in formal notations as a modelling language or semi-formal (as quality procedures).

When modelling a HH process, one has to include the specifications of: the tasks/activities along with their execution order (i.e. routing or control flow), the roles and relations between them and other artefacts concerning organisational issues (e.g. responsibility, availability), the resources ranging from human to devices, the data (that may be "control data" introduced solely for routing purposes or "information data" such as the patient file).

*Home healthcare process model enactment/management:*
Enactment means the manual or automatic execution of an instantiated process model. This domain concerns what has to be provided (process engines, applications or methods) to support the home healthcare process governed by process models; in other words, to provide assistance and guidance for home healthcare activities according to the process specification.

As already mentioned, in this paper, we adopt the term *HH support system* to refer to modelling and enactment support. This system may provide different supports depending on the desired "degree of automation". As we have already mentioned, the home healthcare may cover activities from basic home healthcare to virtual home healthcare, "*HH support system*s" may therefore provide the suitable support for each situation.

The main idea implied by this framework is that for providing relevant and useful support to the home healthcare activities, the process model enactment and the real process domain have to be coupled. If the gap between them becomes important the process enactment will no long influence the ongoing process and the *HH support system* will be rejected by users.

This thesis (that was originally invoked for the software production process) is clearly too restrictive for the home healthcare domain. Indeed unforeseen situations are an intrinsic feature of this domain and process support in this case may provide mechanisms for tolerating deviations between the process and its enactment and must efficiently support the real process even if it only has a partial view of this process. This view depends entirely on the process actor's feedback. Thus, the system is only aware of the activities performed under its control.

*Requirement 1: HH process modelling and enactment*

HH process complexity leads us to this first requirement, untitled *process modelling and enactment*.

Recent investigations in some home healthcare services and organisations pointed out that processes are described in an informal manner, giving guidelines and advices (Arbaoui et al. 2007). However, as related before, the HH process is a complex entity that deals with a lot of constraints (organizational, functional, informational, etc.) and involves several interacting human actors. The process has to become more clearly understood in order to improve communication among people.

An unambiguous representation of the process may provide to the medical coordinator (doctor or nurse) the mean to react to the deviations and to trace the ongoing process. Home healthcare processes need to be clearly represented both for computer execution and user understanding (e.g. medical staff training).

Moreover, medical professionals will be involved in designing HH process models, so modelling languages must be understandable by a variety of users from different domains, not only IT, but also medical professionals.

Therefore a *HH support system* must provide a formal process modelling language with a specific syntax and semantics as well as an enactment support.

HH process models enactment results in automating the activities that can be carried out without the intervention of the human actors (such as sending an alarm signal triggered when some parameters reach a given value) and supporting people in performing the activities that require their participation (like elaborating a "care program" for a newly admitted patient).

Enactment mechanisms have to provide flexible control flow support rather than prescribing a fixed activities ordering. In order to facilitate HH process models communicability the necessity of the enactment of the required formal language should not be made at the expense of their clarity and legibility.

*Requirement 2: HH process evolution*

The second requirement, untitled *process evolution* concerns the *HH support system* ability to support process dynamics and agility, i.e. to respond effectively to dynamic changes.

Dynamic change means that changes are allowed at any time during the HH process model enactment (or on-the-fly). Changes may refer to either unanticipated events resulting from an incomplete specification (at modelling time some aspects of the process may be unknown or too complex), or to handling modifications of the HH process model because of changing conditions. Changes may be induced by several factors, as anticipated events from outside the *HH support system* (e.g. new law, new therapy) or from inside the system (e.g. logical modelling errors, technical problems).

Another critical challenge for a *HH support system* is its ability to manage consistency between real process and process enactment. Rightly, during enactment, the *HH support system* is aware of changes thanks to real process feedback (see figure 1). This feedback allows the enactment domain to gain knowledge in order to make its "image of the reality" as consistent as possible with the state of the real process and therefore makes it more flexible. When the feedback about occurred changes are correctly provided, two possible situations may occur: shared and private evolution.

Shared evolution will affect the enacted process model and therefore will concern current instance (work in progress) and future process instances. This may be the case when, the change is the result of a new strategic organisation and a reengineering effort.

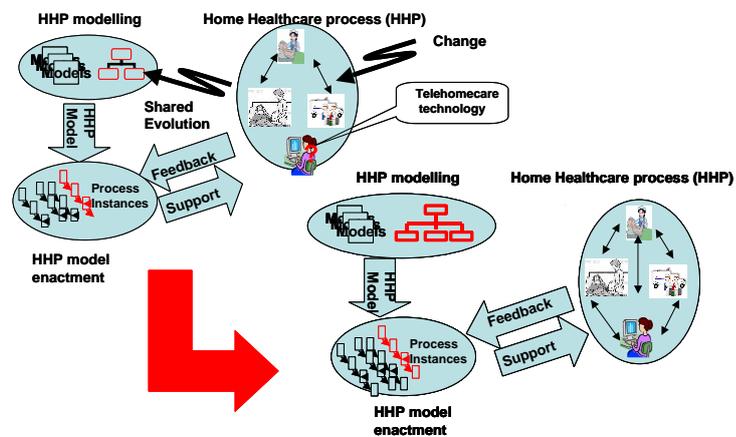

**Figure 2: Shared Evolution**

Figure 2 illustrates this case, let us consider a home healthcare organisation X that uses such *HH support system* and whose mission consists of basic home healthcare (visiting patients, without telesurveillance). If, X decides to use telehomecare technologies, it has to reengineer its process and to move towards another organisation (see on figure 2 two different process models before and after introducing the telehomecare technology). Shared evolution may also be forced by change of Law. This was the case in France when the weekly work time moved from 39 to 35 hours.

Private evolution is used to face unforeseen and exceptional situations that don't necessitate the process model modification but only the current on going instance. In this case the process model is not changed; change will be local to the process model instance that is currently enacted. Here, the change is the result of an error, an exception, or a specific demand of a patient. It is not necessary to change the HH process model, since the change will probably not happen in this form again.

Typical examples, that we have observed in real case of home healthcare context, may be the need to skip an activity (e.g. a patient home visit, step *C* on the figure 3) in the case of an emergency, or the impossibility to provide the care to a given patient that was sent to hospital without informing the HH organisation (due to communication problems between the different partners). As illustrated by the figure 3 the process model is unchanged and the step *C* concerned by the change is bypassed only in the ongoing instance.

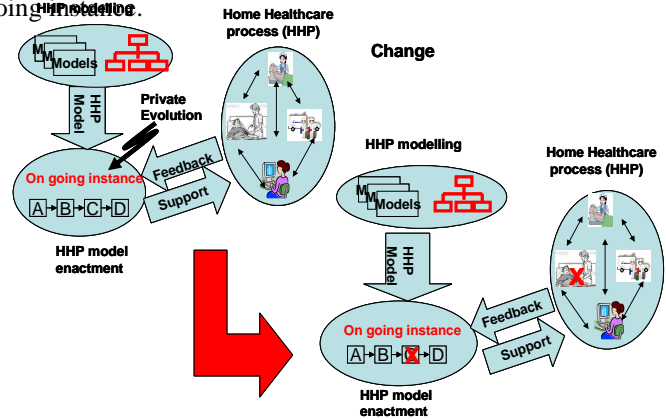

**Figure 3: Private Evolution**

Besides supporting shared and private evolution, answering to process dynamics requires a solution for guarantee that an ongoing instance is compliant with a changed process model. A *HH support system* with such capabilities will be able to manage consistency between the real HH process, the process model and its enactment (see HH process framework in subsection 3-1).

Thus, supporting process evolution is a critical challenge. A *HH support system* that don't meet this requirement may be frequently bypassed by coordinators, nurses, and other HH process participants and be finally considered as an obsolete support.

**4. Home Healthcare support system: Which tracks of solutions?**

As stated before, satisfying these two requirements should be part of the most critical challenges in the HH research domain. Thus we have investigated in 3 process research communities in order to firstly attempt to benefit from past experiences, to highlight strengths along process modelling, enactment and evolution and to derive some open issues.

*HH process modelling and enactment*

Process modelling is a topic that was addressed differently regarding the fixed objectives of the 3 research communities. These objectives ranged from process analysing, improving, simulating, re-engineering to process enactment.

In the *Software process domain,* several Process Modelling Languages (PMLs) have been proposed during the last decade. PMLs are based on a variety of paradigms: logic-based, procedural, production rules, multi-agents, Petri nets, object-oriented, active-databases (Triggers) (Arbaoui et al. 2002), and a great parts are based on multi-paradigms approaches. These PMLs used formal concepts and the resulting process models were enacted within Process Centred Environments (PCEs) in order to support software development. The main benefits provided by these research efforts are powerful process modelling languages that are enactable and that may provide different control flow supports (from prescriptive to proscriptive). Therefore for satisfying this first requirement (i.e. modelling for enacting) this provision of languages and process environments may be useful for developing a *HH support system*. However, little experiments exist in this direction and it may be interesting to see the development of further similar experiments in the future.

In the *Manufacturing process domain*, different approaches have been proposed (Vernadat, 1999) where enactment is not the main issue. These approaches address modelling SADT-IDEF0 (Kim et al. 2003), IDEF3 (Mayer et al. 1995), ARIS (IDS Sheer 2001), ALIX (Pichel et al. 2003), ASCI (Gourgand and Kellert 1991]), architecture modelling CIMOSA (Vernadat 1999), GERAM (Geram 1999), information systems modelling and decision system modelling GRAI, GRAI/GIM (Doumeingts, 1984).
Unlike software process domain, several research works concerning the application of manufacturing process methods and technologies on other processes were conducted and especially on healthcare processes. Among the results of these applications, we can cite: (1) Change management project using company modelling tools (based on GRAI and ARIS), such as transferring different technical platforms towards a partnership structure (Besombes et al. 2004, Trilling et al 2004), (2) Modelling different home healthcare services using IDEF0 in order to analyse and compare their processes (Chahed et al. 2007).
The main benefits provided by these research efforts were the creation of new tools/methods for modelling processes in order to simulate, improve and monitor them. Therefore they represent essential decision making supports for designing a *HH support system*. However, as enactment is not handled, these tools are mostly appropriate upstream and downstream but they are not fit to guide and assist the day to day enactment of the organisational care process.

In the *Business process domain* a variety of approaches to specify and execute workflows have been proposed, based on different workflow languages, as well as tools aiming at supporting business processes modelling and coordination (See website of the Workflow Management Coalition). In a workflow management system, a workflow model is first created to specify organisational business processes and then workflow instances are created to carry out the actual activities described in the workflow model. During execution, the workflow instances can access legacy systems, databases, applications and can interact with users (Georgakopoulos et al. 1995).

The main benefits provided by these research efforts are workflow systems that have been successfully installed and deployed in a wide spectrum of organizations. The same workflow infrastructure may be deployed in various domains (see METEOR (METEOR, 2002)), such as: bio informatics, healthcare, telecommunications, military etc. These systems support workflows involving humans (e.g. coordinating a home healthcare process (Ardissono et al. 2006)), workflows involving systems and applications (e.g. B2B applications), transactional workflows (e.g. a trading process for updating customer orders and financial databases in response to commercial transaction). This provision of modelling, executing, integrating heterogeneous applications and coordinating capabilities, may be useful for developing *HH support systems*.

*HH process evolution*

In the *software process domain*, the major part of Process Centred Environments (PCEs) provide an on-the-fly shared process evolution support but with some extent. The process fragment, or the process activity, concerned by the change can be modified only before its enactment. Different solutions were proposed: such as a set of operators to add, delete activities, specific classes that add new classes or change the classes behaviour during the process model enactment (for a PCE that uses object oriented process modelling language), to delay binding of activities at their enaction time in order to have the possibility to modify them (instead of binding all the activities before the whole process model enactment). None of the studied PCE did provide solutions for modifying a process fragment during its enactment. Indeed, this solution seems technically very complicated to realize as it suppose to stop the fragment enactment, to modify it and then to re-enact it (Arbaoui et al. 2002).

As mentioned before incompleteness of HH process models before their enaction may result in dynamic change situations. This problem was tackled by the software process communities. Indeed two of the developed PCE (academic prototype), provide capabilities to represent uncertain and incomplete knowledge and the appropriate mechanisms for enacting incomplete models (using nonmonotonic logic based AI formalisms) (Arbaoui and Oquendo 1994). Even if these PCEs remain in a prototype state, they provide various "tracks" of research. It seems interesting to study in more details the application of some AI solutions capabilities, in terms of incomplete knowledge representation and nonmonotonic reasoning, for adding flexibility to *HH support systems*.

In the *manufacturing process domain*, the evolution problem is not directly addressed as process model enactment is not the main objective. This issue could be tackled:
- In a dynamic schedule solving problem in accordance with the physical environment constraint. This approach allows the modification of the initial schedule based on information coming from the physical part (Jain and Meeran 1998). Dynamic scheduling rules are also used inside flexible manufacturing systems that was developed in order to propose highly customized product and to respond to volatile demands (Dusonchet 2003).
- Relying on workflow approaches that define different ways for evolution (See Bellow).

In the *business process domain*, the need for flexibility in workflows and the problem of unanticipated exceptions are well recognized by the business process research community. A number of solutions have been proposed. Some of them were inspired from the software process domain (this domain was precursor), at least on the conceptual aspects. Research results concerning the workflow evolution topic may be dispatched in two thematics:
- Managing exceptions in workflow systems (Borgida and Murata 1998, Luo 1998): In the context of workflow management, exceptions are considered to be unexpected undesirable events. In most case exceptions result from applications failure. Tasks which fail, return an exception which is handled by an exception handler (handling an exception consists of either compensate and restart or rollback).

- Managing workflow change (or workflow adaptability) (Van Der Aalst and Jablonski 2000; Riderle et al. 2004): *workflow change* refers to the problem of handling private or shared process evolution and how to migrate existing instances to new workflow definitions. The reason of the change is not necessarily a failure; it may be for example a required modification for improving the process.

The main relevant observation is the multiplicity of published experiments (very recent publications) that use these results in the healthcare domain. This is normal as workflow management systems are already used in order to computerize healthcare environments and are integrated with other healthcare information systems (e.g. electronic patient record systems, financial systems, drug ordering systems, etc.) (Murray 2002, Mangan and Sadiq 2002). Examples of such experiments concern: exception management (Han et al. 2006, Quaglini 2000), but also dynamic change and workflow adaptability support (Brown et al. 2003; Reichert et. al. 2003).

As in the software process domain, the main crucial open issue that is often mentioned concerning workflow adaptability support remains the on-the-fly workflow change and especially the consequences on the ongoing process instances in terms of syntactical and semantic correctness (see model/instances compliance). Until now these consequences were well categorized (Van Der Aalst and Jablonski

2000), the less complicated were implemented (as for modifying a process fragment before its enaction) and the others are still open issues (Reichert et al. 2003).

*An outline of solution tracks*

The investigations of the proposals of these communities regarding process modelling/enactment and evolution lead us to a first idea for combining the best features of these approaches in order to design an effective *HH support system*. As related in the introduction section, the role of a *HH support system* is to support *ICT-home healthcare project*s by providing capabilities for analyzing, modelling, simulating, enacting[1] and improving complex and dynamic HH processes.
The benefits of the manufacturing process domain in such projects may be models/tools for simulating the target solution (upstream) and methods/tools for evaluating and improving the enacted processes. However our study essentially focuses on the contributions that tackle the two requirements, i.e. HH process modelling and enactment and HH process evolution.

Concerning HH process modelling and enactment:
- The benefits of software process approaches are: goal oriented modelling and enaction (maybe an answer for dynamic ordering), underlying formalisms such as: logic (for temporal constraints description, non monotonic for uncertain knowledge representation), AI (e.g. multi-agents for modelling cooperation, negociation ). The downside of these approaches is that very few experiments have been carried out in other domains.
- The benefits of business process approaches are: integration and interoperability solution (essential for home healthcare in regard to heterogeneous partners, use of legacy system, hospital information systems), graphical interfaces that facilitate process models communication. The downsides of these approaches are that today workflow management systems only support syntactical checks (Van der Aalst 2003). Verification and logical correctness of workflow definition is limited, modelling languages need in this case more theoretical foundations.

Concerning HH process evolution:
Both software process and workflow technologies tackled the process evolution problem and are still facing on-the-fly change problems.
In the software process domain we can learn much about formal support (formal solution) in response to this problem, like modal logics for instance. However, workflow management systems provide more commercialized well-tried tools, with a very attractive graphical user interface. This criteria doesn't solve the dynamic change problem, commercial workflow management systems do not allow change propagation to on-going process instances (Rinderle et al. 2004). So, the challenge to effectively support evolution may consist in successfully combining the best features of these approaches:

- Theoretical features from the software process side regarding the expressiveness as well as the formal and operational semantics of software process modelling languages,
- technological and organisational features from the business process side regarding capabilities inherited from the information systems domain e.g. schema evolution, transaction management (for exception handling), those acquired from first experiments in several HH domains and especially in the home healthcare (Ardissono et al. 2006).

**5. Conclusion**

In this paper, our focus is laid on the home healthcare process, considered as one of the "human intensive processes". We advocate that the mastering of HH process is a necessary condition to provide an efficient support to HH activities. We introduce the concept of *HH support system* and highlight the requirements that have to be met by such systems in order to deal with HH process complexity and dynamics.

Obviously, we haven't tackled every aspect of HH processes. Features such as distribution/decentralization are worth exploring. HH processes are highly distributed (distribution of responsibilities, technologies, resources ). Distribution has social and technical aspects. A *HH support system* must enable cooperation between different actors (mobile actors) and between heterogeneous systems. A comprehensive answer to home healthcare activities must absolutely integrate the 2 aspects discussed in this paper (process complexity and dynamics) plus the decentralization and its relation to them.

Afterwards, we present how the selected requirements have been tackled in three process communities: software process, manufacturing process, business process. This leads us on the one hand to a proposition outline on how to combine these technologies for the development of a *HH support system* and on the another hand to open issues that still need research investigations:

- How to reconcile multiple paradigms in order to get an accurate language with rigorous syntax and semantics, with legibility and ease of use (to cope with process complexity, heterogeneous sub-services and multi-disciplinarily actors).
- How to manage incomplete and uncertain knowledge (to cope with dynamic change due to the enactment of incomplete models).
- How to propagate change on current and future process instances while satisfying compliance constraints between the state of these instances and the state of the process model (to cope with long running HH processes that are frequently adapted).
- How to involve HH process actors in the management of dynamic change, either a momentary (exception) or permanent change (to cope with heterogeneous and multi-disciplinary actors).

We can conclude by attesting that developing "computerized" support for HH processes is a challenging task that requires different skills from disciplines such as:

software engineering, industrial and business engineering, information systems. This is a direct consequence of the complexity of such highly human processes.

Future works will deal more specifically with HH process evolution along with supporting incomplete and uncertain knowledge management. The main challenge will be to incorporate more formal concepts in order to support dynamic change process, without neglecting the usability and legibility of the modelling formalism. We are still in the first step of this research work, that consists of going more in depth in the way of working of HH organisations by observing, modelling and analysing processes and especially dynamic changes situations, trying to answer questions such as: which kind of changes? How they are handled? Etc.